\documentclass[apl,aps,preprint,showpacs]{revtex4}
\usepackage{amssymb}

\usepackage{graphicx}

\usepackage[T1]{fontenc}

\usepackage[american]{babel}

\begin{document}

\title{Mesoscopic electronic heterogeneities in the transport properties of V$_2$O$_3$ thin films}

\author{C.~Grygiel}
\affiliation{Laboratoire CRISMAT, UMR 6508 du CNRS, ENSICAEN et Universit$\acute{e}$ de Caen, 6 Bd Mar$\acute{e}$chal Juin, F-14050 Caen 4.} 
\author{A.~Pautrat}
\affiliation{Laboratoire CRISMAT, UMR 6508 du CNRS, ENSICAEN et Universit$\acute{e}$ de Caen, 6 Bd Mar$\acute{e}$chal Juin, F-14050 Caen 4.} 
\author{W.~C.~Sheets}
\affiliation{Laboratoire CRISMAT, UMR 6508 du CNRS, ENSICAEN et Universit$\acute{e}$ de Caen, 6 Bd Mar$\acute{e}$chal Juin, F-14050 Caen 4.}
\author{W.~Prellier}
\affiliation{Laboratoire CRISMAT, UMR 6508 du CNRS, ENSICAEN et Universit$\acute{e}$ de Caen, 6 Bd Mar$\acute{e}$chal Juin, F-14050 Caen 4.} 
\author{B.~Mercey}
\affiliation{Laboratoire CRISMAT, UMR 6508 du CNRS, ENSICAEN et Universit$\acute{e}$ de Caen, 6 Bd Mar$\acute{e}$chal Juin, F-14050 Caen 4.} 
\author{L.~M\'echin}
\affiliation{GREYC, UMR 6072 du CNRS, ENSICAEN et Universit$\acute{e}$ de Caen, 6 Bd Mar$\acute{e}$chal Juin, F-14050 Caen 4.}

\date{\today}

\vspace{8.5cm}

\begin{abstract}

The spectacular metal-to-insulator transition of V$_2$O$_3$ can be progressively suppressed in thin film samples. 
Evidence for phase separation was observed using microbridges as a mesoscopic probe of transport properties where 
the same film possesses domains that exhibit a metal-to-insulator transition with clear first order features or remain metallic down to low temperatures. 
A simple model consisting of two parallel resistors can be used to quantify a phase coexistence scenario explaining the measured macroscopic transport properties. 
The interaction between film and substrate is the most plausible candidate to explain this extended phase coexistence as shown by a correlation between the transport properties
 and the structural data.

\end{abstract}


\maketitle

\newpage
\quad 

Epitaxial thin films have been the focus of intensive research, owing to their potential application in emergent technologies and fundamental studies of physical 
phenomena. In particular, strain introduced by a lattice mismatch between thin film and substrate, can be used to alter their properties 
and stabilize metastable states. An extended temperature regime for phase coexistence has been observed for certain chemical compositions that, in principle, 
are not allowed by the Gibbs' phase rule.\cite{gibbs} One well documented case is that of perovskite manganite R$_{1-x}$A$_x$MnO$_3$ ($R$ = rare earth 
and $A$ = alkaline earth cations)
thin films, which undergo a metal-to-insulator phase transition at a significantly different temperature than those of bulk samples.\cite{manga} It should also be 
noted that the 
disorder in the ionic radius of the $A$-site cation has a profound effect on the average ordering temperature of bulk manganites, which often complicates the determination of how much strain 
contributes to phase separation in thin film samples.\cite{moreo1999,dagotto2001} Nevertheless, a recent comparison of solid-solution alloy and $A$-site ordered 
superlattice La$_{2/3}$Ca$_{1/3}$MnO$_3$ thin 
film samples on different substrates revealed that strain is more important than chemical disorder in stabilizing the mixed phase regime near the average ordering 
temperature.\cite{pala1} To investigate further the role of the substrate in the phase separation of a thin film sample,
 it is interesting to work on a sample which is, in its bulk form, a prototype of a discontinuous transition.
 In this contribution, the transport properties of epitaxial V$_2$O$_3$ thin films are examined. 

Vanadium sesquioxide V$_2$O$_3$ exhibits a remarkable first-order metal-to-insulator (M-I) transition around 150\,{K}, below which an antiferromagnetic insulating phase 
exists. Hydrostatic and chemical pressure (V$_{2-x}$M$_{x}$O$_3$ with M=Ti...) or sample non-stoichiometry (V$_{2-y}$O$_3$) significantly lowers this 
transition temperature.\cite{shivashankar83,ueda80,whan73,yethiraj90} In particular, application of hydrostatic pressure above a critical threshold of 26\,{KBars} suppresses 
the M-I phase transition.\cite{whan69} Under particular growth conditions, substrate-induced strain also suppresses partially the M-I transition,\cite{lab1,allimi} 
and metallic-like behavior, with some indication of a Fermi liquid regime, is observed below 20\,{K}.\cite{lab1} The intermediate-temperature regime, which is apparently dependent upon film 
thickness, is more complex, and a non-monotonic variation of the slope ($dR$/$dT$) is observed.\cite{grygiel2007} Surprisingly, 
using c-Al$_2$O$_3$ \cite{grygiel2007} or c-LiTaO$_3$ \cite{allimi} , the thicker films present a metallic character. 
Films thinner than 220\,{\AA} recover the M-I transition, although it is attenuated strongly when compared to the transition observed in the bulk. 
To clarify such behavior, 
we have investigated the homogeneity of the samples at the mesoscopic level (where mesoscopic represents a scale below the large statistical collections at which average properties 
have been measured previously). 
Using microbridges as local probes, we show in these thin films a phase separation, which explains by using a simple model of two parallel resistors 
the evolution of the transport properties at the macroscopic scale. A clear correlation with the structural data is also found.

\begin{figure}[htb!]
\begin{center}
\includegraphics*[width=9cm]{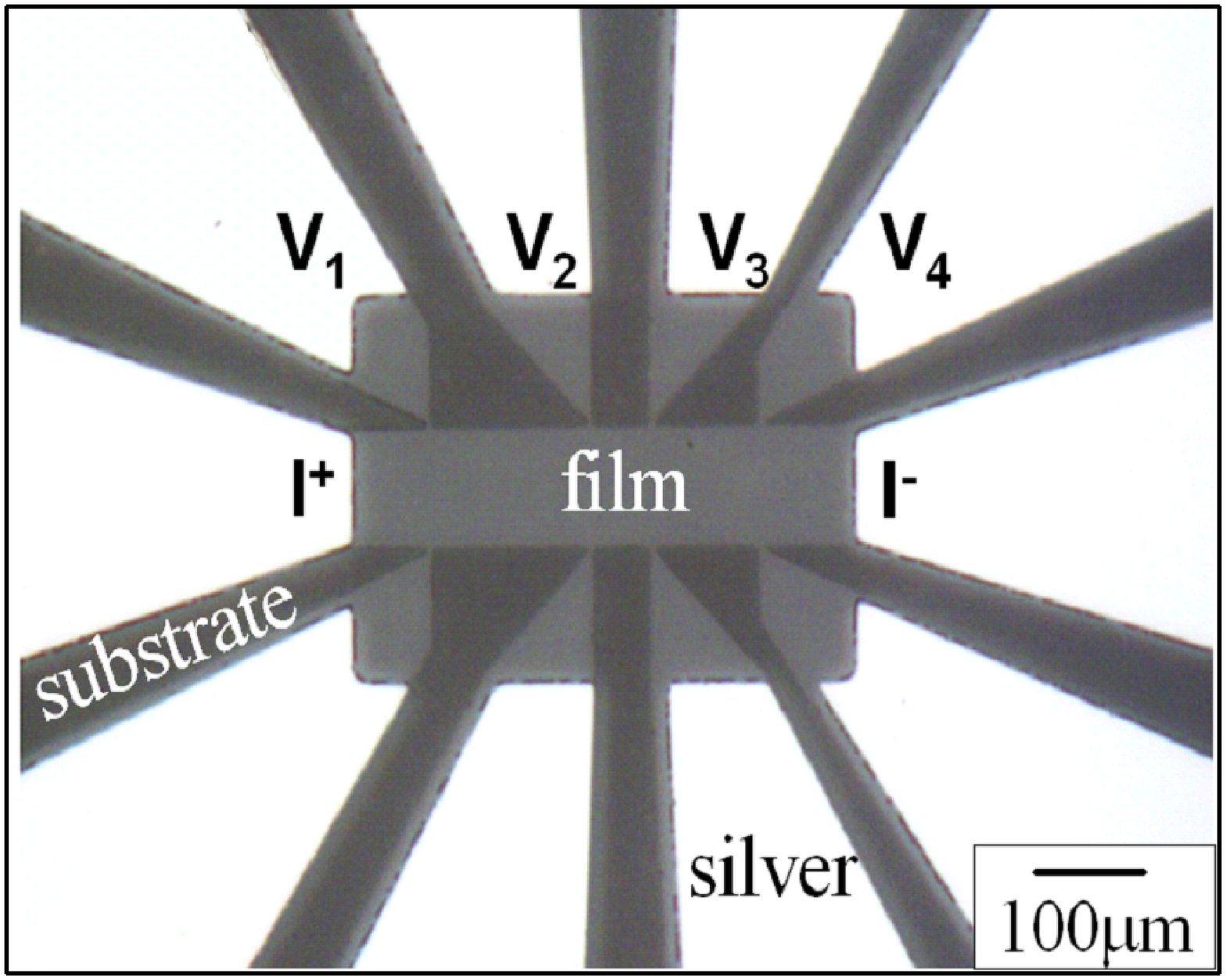}
\end{center}
\caption{An optical microscopy image of one microbridge patterned (M2, width : 100\,{$\mu m$}) for a 230\,{\AA} thick V$_2$O$_3$ film, where the scale 
represents 100\,{$\mu m$}. The microbridge includes two silver pads for supplying the current ($I^{+}$, $I^{-}$) and eight silver pads with five different lengths 
depending on the position of the voltage contacts ($V_1$, $V_2$, $V_3$ and $V_4$). Our measurements were made between $V_1$ and $V_3$, a distance corresponding 
to 200\,{$\mu m$}.}
\label{f1}
\end{figure}

High quality V$_2$O$_3$ films were grown on (0001)-oriented sapphire substrates 
using the pulsed laser deposition technique. A pulsed KrF excimer laser beam ($\lambda$=248\,{nm}, pulse length 20\,{ns}, repetition rate 3\,{Hz}) was focused on a V$_2$O$_5$ target, depositing 
films onto the substrate under optimized deposition conditions (650${{}^{\circ }}$C, 0.02\,{mbar} Ar pressure,
and a laser fluence of 4\,{J cm$^{-2}$}).
The structure of the resulting films was examined by X-Ray Diffraction, which attests to their high quality and epitaxial relation with the substrate. The in-depth 
details of the growth conditions, and some structural and microstructural properties have been reported previously.\cite{grygiel2007} 
The resistance of the samples was measured using a physical property measurement system. 
Electrical transport measurements were made on unpatterned films using a four probe geometry. Silver contact pads, separated by 1\,{mm}, were deposited 
by thermal evaporation through a mechanical mask. A 230\,{\AA}
thick sample, which exhibits a low rms roughness of 0.47\,{nm}
(averaged over 3$\times$3$\mu m^{2}$), was selected for patterning.
A silver layer was deposited onto the film. Contact pads were first
defined using ultraviolet (UV) photolithography and chemical etching
in a KI/I$_{\text{2}}$ solution. The V$_{2}$O$_{3}$ thin film
was then patterned by UV photolithography and argon ion etched to
form 20\,{$\mu $m} and 100\,{$\mu $m} wide bridges. The final microbridge, shown
in the figure \ref{f1}, allows for different measurement geometries. For this experiment, the voltage was measured between $V_1$ and $V_3$, which 
corresponds to a length of 200\,{$\mu $m}. 
The  external circuitry
and the patterned film with silver contact pads were connected
using aluminum-silicon wires attached by ultrasonic bonding. Transport measurements
were made in 20$\times$200$\mu m^{2}$ (M1) and 100$\times$200$\mu m^{2}$
(M2) bridges. M1 and M2 are located close to each other in the center of the sample. In
unpatterned films, we find that the resistivity has an average value of $450 \pm 150 $\,{$\mu \Omega $.cm} at room temperature.
 In patterned film, we measure 1030$\pm $80\,{$\mu \Omega $.cm} for M1 and 950$\pm $80\,{$\mu \Omega $.cm} for M2. This increase is almost entirely due to an increase of the residual resistivity. From this we conclude that the patterning process did
not significantly alter the film properties and, importantly, that each bridge is nearly identical.

\begin{figure}[htb!]
\begin{center}
\includegraphics*[width=9cm]{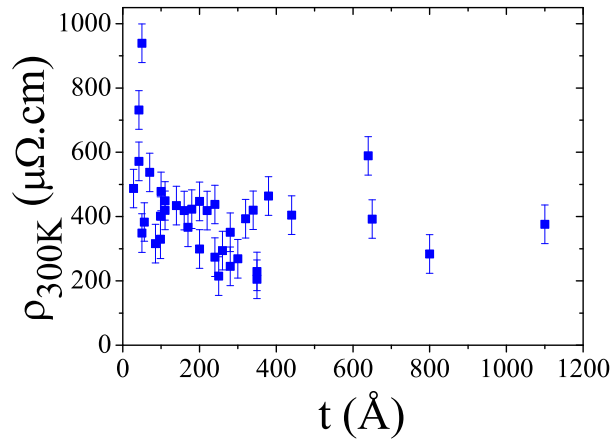}
\end{center}
\caption{Room temperature resistivity values ($\rho _{300K}$) versus the thickness ($t$) of unpatterned V$_2$O$_3$ films.}
\label{rho300}
\end{figure}

\begin{figure}[htb!]
\begin{center}
\includegraphics*[width=16cm]{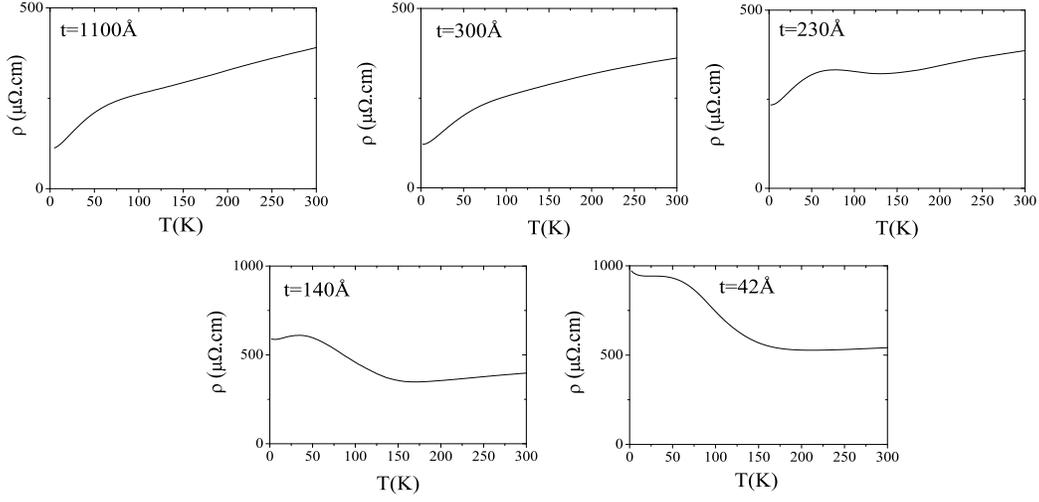}
\end{center}
\caption{Linear resistivity as function of the temperature measured during cooling for unpatterned V$_2$O$_3$ films of certain thicknesses (noted $t$).}
\label{courbes-ech}
\end{figure}

\begin{figure}[htb!]
\begin{center}
\includegraphics*[width=9cm]{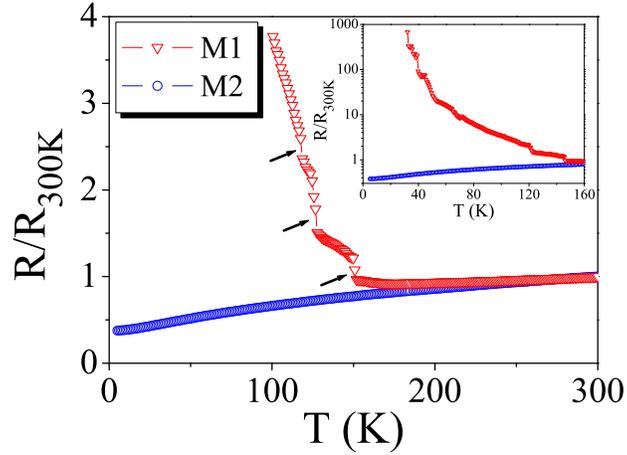}
\end{center}
\caption{The resistance, normalized over the 300K value, of a patterned 230\,{\AA} thin film of V$_2$O$_3$ measured during cooling on two 
different microbridges (M1 : 20$\times$200$\mu m^2$, M2 : 100$\times$200$\mu m^2$). The arrows represent for M1 different transition temperatures observed. 
The inset is a close-up near the M-I transition and down to low temperatures.}
\label{f3}
\end{figure}

The resistivity of numerous unpatterned V$_2$O$_3$ thin films were measured as a function of their thickness 
(approximately 20 samples with thicknesses $t$, 40\,{\AA}$\leq t\leq$ 1000\,{\AA}).
As shown in figure \ref{rho300}, the film resistivities at room temperature are close to $450 \pm 150 $\,{$\mu \Omega $.cm},
values which are similar to the those of the crystals,
 and do not evolve with the thickness. The resistivity versus temperature curves 
measured during cooling for samples of different thicknesses are presented in figure \ref{courbes-ech}. 
While the high temperature behavior is similar for all the samples notable differences appear when $T\leq 150$\,{K}. 
Insulating behavior is favored for the thinnest samples and metallic for the thickest. A pure metallic behavior was never observed 
at the macroscopic scale over the 2-300\,{K} range 
since a memory of the M-I transition is always observed (even if it is tiny), but metallic behavior is recovered at low temperatures.\cite{grygiel2007} 
It should be noted that even the more metallic V$_2$O$_3$ behaves as a bad metal
 and the classical size effect is not observed down to the lowest thickness ($t \approx 42$\,{\AA}), which is 
in agreement with the anomalously low mean free path ($\ell \leq 0.2nm $ in a Drude approximation). This explains the thickness independence of the room temperature resistivity.
 At lower temperatures, the complex behavior of the resistivity requires further
 investigations. In particular, the resistivity maximum occuring at intermediate temperature may indicate a competition between metallic and insulating 
states as observed previously in manganites.\cite{mayr} 
The phase separation scenario has been used to explain such characteristics and one may ask if the situation is similar in our films.  
Accordingly, we have performed local resistivity measurements on a sample possessing an intermediate thickness ($t$=230\,{\AA}). This film was 
selected because it represents the cross-over between the metallic and the insulating unpatterned film behaviours, and displays a clear maximum in $\rho(T)$ at $T\approx 70K$.  
Local resistivity measurements were done using two microbridges (M1 and M2) patterned on this sample in order to observe if the transport properties are identical at a smaller scale than the macroscopic one. 
Figure \ref{f3} displays the evolution of the resistance ratio $R$/$R_{300K}$ for each microbridge of this patterned film. From the plots, it is
 clear that each microbridge exhibits significantly different transport properties for $T\leq 150$\,{K}, and that the macroscopic behavior is not simply recovered.

\begin{figure}[htb!]
\begin{center}
\includegraphics*[width=9cm]{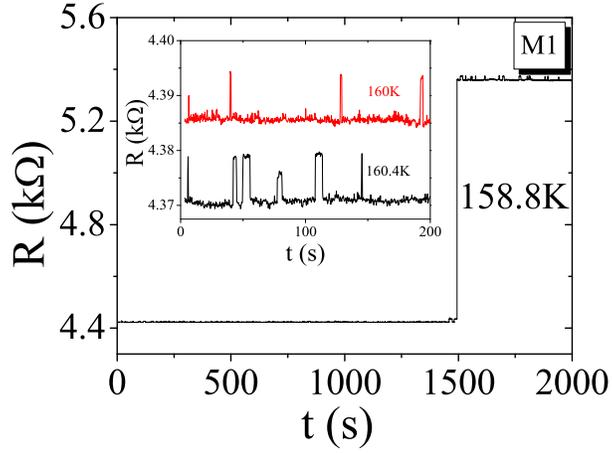}
\end{center}
\caption{Resistance time series after supercooling for the insulating microbridge M1 of 230\,{\AA} thick V$_2$O$_3$ film at 158.8\,{K}. 
The inset shows the resistance fluctuations of two states respectively at 160.4\,{K} and 160\,{K}.}
\label{noise}
\end{figure}

The data for the first microbridge (M1) present a metal-to-insulator transition. 
While the largest increase in resistance is observed close to 150\,{K}, smaller ones are also observed at certain discrete values: 120\,{K}, 110\,{K} 
and at approximately each 10\,{K} down to 70\,{K} (as shown on figure \ref{f3} by arrows).
 The transition at 150\,{K} appears similar to the first order M-I
transition observed in pure bulk V$_2$O$_3$ samples. The resistance increases around three orders across the transition between 150\,{K} and the low temperatures, and
 this is notably less than what is reported for the transition in crystals 
where the increase can be up to seven orders of magnitude. Since other transitions occur at different temperatures, each jump in resistance 
can be attributed to the transition of a part of, and not all of, the bridge. Such behavior can explain, at least partially, why the M-I transition is notably attenuated.
Owing to the multiple transitions observed, the domain size of one insulating domain is estimated to be smaller than that of the microbridge (< 20$\times$200$\mu m^{2}$),
 but not too small to largely affect the resistivity, i.e. in the micrometer scale. 
 Two features demonstrate that this transition remains locally first order,
 even though it is affected by the disorder. Firstly, after cooling the sample, resistance time series were measured at fixed temperatures (Fig.~\ref{noise}). 
During these measurements resistance fluctuations were observed, which are large non-Gaussian $1/f^{\alpha}$ noise, and can be attributed to highly inhomogeneous current
 paths that form when close to an incipient M-I transition,\cite{DCR} or two states fluctuators.
 We focus here on the discrete two-state fluctuators near the transition, shown in the inset of figure \ref{noise}, in the absence 
of large non-equilibrium noise. 
The statistics of the two states are largely independent of the applied magnetic field, which demonstrates that
 the fluctuation of magnetic domains can not be involved as an origin of this noise (the M-I transition is also a paramagnetic to antiferromagnetic transition, so magnetic effects can 
not be completely neglected). 
Based on work in colossal magnetoresistive films,\cite{merithew} the measurement of the resistance during long intervals (approximately 1 hour for each 
time series) can be used to extract the main relaxation time $\tau_i$ for each state (high resistance (i=1) and low resistance (i=2) states). 
Assuming $\tau_i \propto exp(-G_i/K_BT)$ where $G_i$ is the free energy in the state $i$, the Boltzman factor $r=\tau_1/\tau_2$ depends on the free energy difference.
 The latter term then can be used to calculate the entropy difference ($\Delta S$), which we find to be $\Delta S \approx  60 K_B$ using the thermodynamic 
identity $\Delta S = K_B \partial(T ln r) /\partial T$. Such behavior is expected when a barrier between two phases exists,
 and the transition is first order (rather than second order).\cite{pala}
Secondly, when supercooling from a high temperature, the 
low resistance state is blocked, and after a certain time the resistance switches suddenly to its high resistance state (see Fig.~\ref{noise}). 
 This indicates that a metastable state can be supercooled, which is also in agreement with a first order nature for the transition.

The second curve on the figure \ref{f3} corresponds to the measurements made on the second microbridge (M2) on the same sample.
For this microbridge there is no trace of the M-I transition and the sample behaves as a metal down 
to 2\,{K}. 
At low temperature, the resistance follows the expected $T^2$ temperature variation for electron-electron scattering, as already shown for V$_2$O$_3$ thin films 
and single crystals subjected to
high pressures (26-52\,{KBars}).\cite{grygiel2007,whan69} The detailed analysis of this metallic phase, including magnetoresistance, noise properties,
and localisation effects will be discussed in another paper. Here, we address if the suppression of the M-I transition in these domains can be accounted for substrate induced pressure.
Assuming the film is confined over the surface area of the substrate at the M-I transition temperature leads to an increase of pressure in the film.
The maximum effective pressure 
generated can be estimated \cite{pala} using experimental data from bulk samples, such as the stress coefficients $C_{11}$
 and $C_{12}$,\cite{stress} and the volume change during the bulk M-I transition
 (we take 2.3 $\%$ as a representative value, even if some dispersion exists in the litterature).\cite{whan69}
 This results in a calculated effective pressure of ($\Delta P$)$\approx 27 KBars$, and because 
the decrease in the critical temperature owing to elastic distortion arises from the experimental $dP$/$dT_c$,\cite{whan69} the variation 
in the critical temperature is ($\Delta T_c$)$ \geq 160 K$. 
This value is consistent with the suppression of the M-I transition. 
Since a first order transition is heterogeneous when the volume is kept constant because it induces an heterogeneous pressure,
such mechanism will certainly lead to a broad distribution of critical temperatures at a macroscopic scale. As a consequence, for a fixed temperature $T<150 K$,
both metallic and insulating domains with different critical temperatures can coexist.

These results obtained for the microbridges indicate the presence of mesoscopic electronic heterogeneity in V$_2$O$_3$ films where metallic and insulating regions coexist over a very large temperature range.
Since their properties contrast, clear consequences can be expected for transport properties measured on a large scale, i.e. for unpatterned samples. 
We suggest that these films consist of different metallic and insulating domains, such as those observed using the microbridges, which are formed below 150\,{K}. 
In heterogeneous medium, percolation theory can be used to calculate the effective resistivity \cite{perco}, using for example random resistors networks.
 In the case of large heterogeneities, a coarse-grain approach can be sufficient to give a good description of the resistivity,
 and the effective resistivity can be described using a parallel-resistor model with a metallic (percolative) resistance and an insulating resistance \cite{mayr}. We do not have a direct measure of the size of the domains, but as discussed above,
 we propose that the each insulating domain is in the micrometer scale. In the metallic microbridge M2,
 we do not observe any features (non linearities in voltage-current characteristics, thermal hysteresis) which could indicate mixed states at a lower scale, showing that metallic domains are homogeneous over the size of the microbridge or that it is always shorted out by metallic paths,
 i.e. above the percolation threshold.
In phase separated manganites, a coarse grain approach was strongly guided by the existence of a peak at intermediate temperatures in the effective resistivity,
 by the large scale (micrometric) of the electronic heterogeneities, and by the recovery to metallic properties at low temperature indicating that current flows trough metallic paths \cite{mayr,andres,taran}.
Here we observe the same characteristics. 
Consequently, to reconstruct the observed macroscopic behaviours, we also propose a two-parallel-resistors model, with each resistance directly given by the measured resistance in the microbridge (M1 for $R_I$ and M2 for $R_M$).
The ratio of the metallic phase, denoted as $x$ can be used in the 
following expression to determine the effective conductivity $\sigma$,
 
\begin{equation}
\label{"the modele"}
\sigma  = x \sigma _M  + (1-x) \sigma _I ,
\end{equation}

where $\sigma _M$ and $\sigma _I$ are the conductivity of the metallic and insulating phases of the film, respectively, measured experimentally with the microbridges.
 The variable in this basic model, except for the $x$ value, is the value of the residual resistivity of the metallic phase,
 which depends slightly (in a non-trivial way) on the fluctuations of the growth parameters. This variable influences the absolute resistivity value
 at low temperatures where the current path is essentially in the metallic phase, but, overall, does not significantly alter the functional form of the $R(T)$.

\begin{figure}[htb!]
\begin{center}
\includegraphics*[width=16cm]{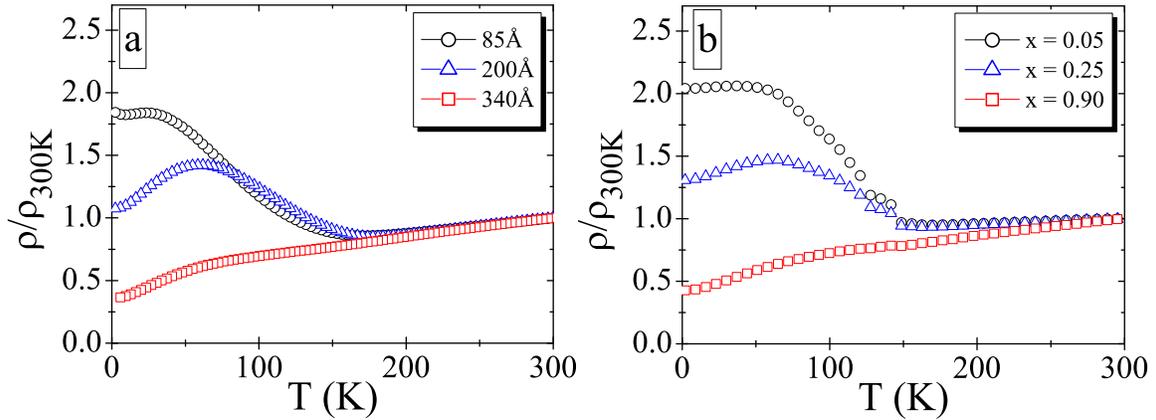}
\end{center}
\caption{a) Normalized resistivity ratio $\rho(T)/\rho(300K)$ as function of temperature measured during cooling for three thicknesses. b) Simulated resistivity ratio ($\rho $/$\rho _{300K}$) versus temperature for the model of parallel resistances applied to certain metallic ratios $x$.}
\label{courbes-modele}
\end{figure}

\begin{figure}[htb!]
\begin{center}
\includegraphics*[width=9cm]{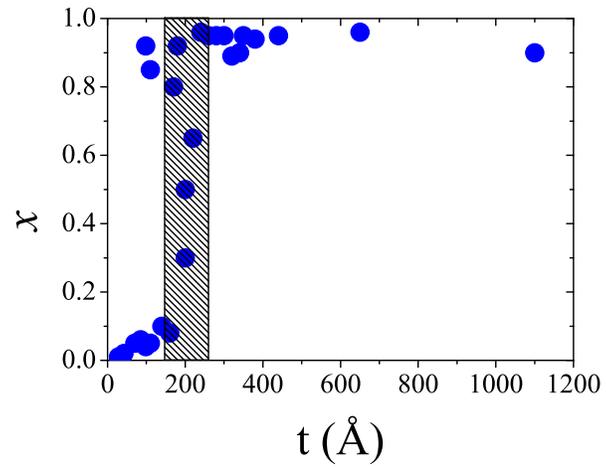}
\end{center}
\caption{Ratio $x$ of the metallic phase, extracted from the application of the parallel-resistance model, versus thickness of our V$_2$O$_3$ films. 
The hatching zone corresponds to the cross-over thickness, which is close to 200\,{\AA}.}
\label{x(t)}
\end{figure}

We have applied this phenomenological model to our unpatterned films. Three simulated curves of resistivity as function of the temperature, each corresponding to a different $x$, are shown 
in the figure \ref{courbes-modele}. The $x$ values, which are temperature independent, were ajusted to match
 the temperature dependence of simulated curves with those of the experimental curves. 
The experimental and the calculated curves have very similar appearances, demonstrating that the model (even if simple) is able to explain semi-quantitatively the transport properties.
 Consequently, the relevant parameter depending on the thickness is the ratio of conducting phase, $x$. 
For each sample, the corresponding $x$ parameter was deduced. 
The figure \ref{x(t)} shows its thickness-dependence, with a large change at a critical thickness $t_c \approx 200$\,{\AA} 
(as shown on the curve by the hatching zone). For $t\geq t_c$, $x$ is close to $1$, indicating that almost the whole film is metallic. 
In contrast, for $t<t_c$, $x$ becomes small and 
the domains with the M-I transition form the majority of sample, with an electrical behaviour similar to the bulk material. 
The film thickness can therefore be used to control 
the ratio of conducting phase and the mesoscopic phase separation. 
The range of thickness where the phase separation is critical is restricted to the domains of $100$\,{\AA}$<t<250$\,{\AA} (see Fig.~\ref{x(t)}).

These observations show that the metallic phase  in our V$_2$O$_3$ films is favored by the large thicknesses. Recalling that metallic V$_2$O$_3$ crystals 
result from an effective applied pressure, we come to the suprising conclusion
 that the thicker samples are under larger stress than the ultra-thin films.
 
\begin{figure}[htb!]
\begin{center}
\includegraphics*[width=9cm]{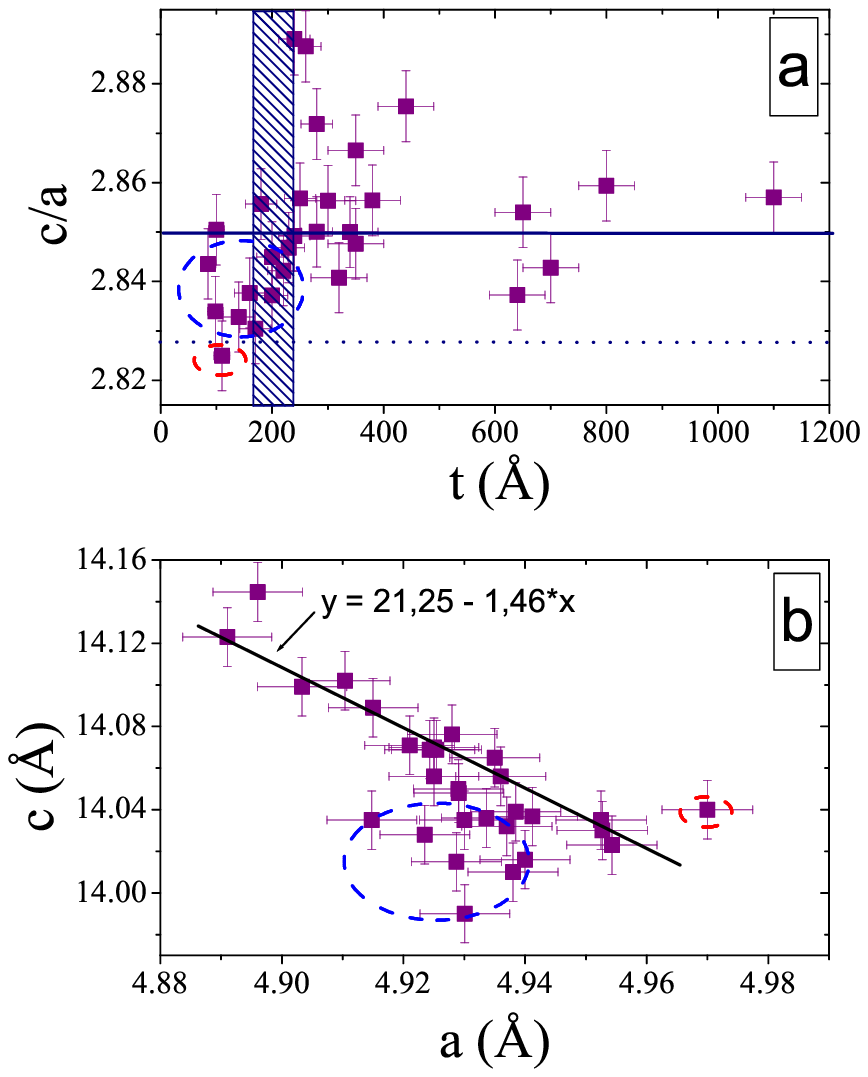}
\end{center}
\caption{a) : $c/a$ ratio of the lattice parameters versus thickness of V$_2$O$_3$ films at room temperature. The hatching zone corresponds to the cross-over thickness, close to 200\,{\AA}. 
The doted line corresponds to the bulk value, whereas the solid line refers to the 
value for crystals subjected to a hydrostatic pressure of 47\,{kbar}.\cite{finger80} 
b) : Out-of-plane lattice parameter ($c$) versus in-plane lattice parameter ($a$) for the different films. The line indicates films following the Poisson law. 
On both curves, the surrounded points correspond to the same sample thickness ($t< 200$\,{\AA}, see the text for details).}
\label{c/a}
\end{figure}

To explain this unusual evolution of the properties, a structural study was performed. 
It is known that the ratio of lattice parameters $c/a$ is extremely sensitive to lattice distorsions, especially in the case of hexagonal 
stacking.\cite{finger80,luo2004} Figure \ref{c/a}a represents the evolution of $c/a$ as a function of the film thickness, deduced from the analysis of X-Ray diffraction at room temperature.\cite{grygiel2007} 
A small variation in this ratio is observed close to a thickness 
of 200\,{\AA} (indicated by a dotted line in figure \ref{c/a}a), which is close to the $t_c$ deduced from the transport properties. 
Films with $t<t_c$ have a reduced ratio $c/a$ ($\sim 2.835$), and compare favorably with the value of  $2.828$ reported for a pure $V_2O_3$ crystal (dashed line). 
Films with $t>t_c$ possess a ratio closer to $2.85$, the value for a crystal subjected to high
hydrostatic pressure (solid line).\cite{finger80} 
Further indication of a structural cross-over at $t\approx t_c$ is observed in figure \ref{c/a}b. The thick films ($t>t_c$) 
follow a Poisson law with a calculated Poisson coefficient $\nu \approx 0.5$ 
(incompressibility limit). In contrast films with $t<t_c$ do not follow the Poisson law, points surrounded on the curves (Fig.\ref{c/a}), indicating a change in the elastic properties of the films. 
Overall, both observations 
indicate that the thinner films have structural properties close to those of the bulk material, and thus undergo the M-I transition associated with a volume variation.

We have shown that heterogeneous strains induced by the M-I phase change and substrate-induced volume confinement are likely responsible for the broadening of the 
critical temperature from 150\,{K} until complete suppression for the conducting phase.
However, the role of non-stoichiometry which is known to influence $V_2O_3$ properties must also be considered.
 For V$_{2-y}$O$_3$ compounds, the ratio $c/a$ remains constant, whatever the $y$ value, and is equal to $2.828$ because 
both $a$ and $c$ are compressed with respect to stoichiometric $V_2O_3$.\cite{ueda80}
 In our films, $c/a$ varies with $t$ and both $a$ and $c$ exhibit an in-plane
compressive lattice parameter and an out-of-plane tensile lattice parameter.\cite{grygiel2007} In general, the evolution of properties as function of $t$
 in our films can not be explained by non-stoichiometry, but is consistent with pressure effects.\cite{ueda80,whan69} 
Further studies, including detailed structural studies with Transmission Electron Microscopy,
 are under way to understand the details of the growth process responsible for the cross-over in the unit cell distortion at $t\approx 200$\,{\AA}.

In summary, we have used microbridges as local probes to examine the transport properties of V$_2$O$_3$ thin films.
Mesoscopic electronic heterogeneities are observed and consist of insulating and metallic zones that coexist over a very large temperature range.
This phase separation has a strong influence on the macroscopic transport properties where the evolution of the resistivity as a function of film thickness appears to be 
governed by the change in quantity of each phase, which is in agreement with the structural data. As in the case of manganite films,\cite{pala} 
we believe that such heterogeneous states
 should be common for films with first order transitions interacting with a substrate.  

This work is carried out in the frame of the STREP CoMePhS 
(NMP3-CT-2005-517039) supported by the European Commission and by the CNRS, France. W.~C.~Sheets was supported additionally by a Chateaubriand postdoctoral 
fellowship.

\newpage

\begin{thebibliography}{99}

\bibitem{gibbs} V.~M.~Kaganer, B.~Jenichen, F.~Schippan, W.~Braun, L.~D$\ddot{a}$weritz and K.~H.~Ploog, Phys. Rev. Lett. \textbf{85}, 341 (2000).

\bibitem{manga} W.~Prellier, P.~Lecoeur and B.~Mercey, J. Phys. Condens. Matter \textbf{13}, R915 (2001).

\bibitem{moreo1999} A.~Moreo, S.~Yunoki and E.~Dagotto, Science \textbf{283}, 2034 (1999).

\bibitem{dagotto2001} E.~Dagotto, T.~Hotta and A.~Moreo, Phys. Rep. \textbf{344}, 1 (2001).

\bibitem{pala1} A.~Palanisami, M.~P.~Warusawithana, J.~N.~Eckstein, M.~B.~Weissman and N.~D.~Mathur, Phys. Rev. B \textbf{72}, 024454 (2005)

\bibitem{shivashankar83}  S.~A.~Shivashankar and J.~M.~Honig, \prb \textbf{28}, 5695 (1983).

\bibitem{ueda80}  Y.~Ueda, K.~Kosuge and S.~Kachi, J. Solid State Chem. \textbf{31}, 171(1980).

\bibitem{whan73}  D.~B.~McWhan, A.~Menth, J.~P.~Remeika, W.~F.~Brinkman and T.~M.~Rice, Phys. Rev. B \textbf{7}, 1920 (1973).

\bibitem{yethiraj90}  M.~Yethiraj, J. Solid State Chem. \textbf{88}, 53 (1990).

\bibitem{whan69}  D.~B.~McWhan and T.~M.~Rice, Phys. Rev. Lett. \textbf{22}, 887 (1969).

\bibitem{lab1} S.~Autier-Laurent, B.~Mercey, D.~Chippaux, P.~Limelette and Ch.~Simon, Phys. Rev. B \textbf{74}, 195109 (2006).

\bibitem{allimi} B. S. Allimi, M. Aindow, and S. P. Alpay, Appl. Phys. Lett. \textbf{93}, 112109 (2008).

\bibitem{grygiel2007} C.~Grygiel, Ch.~Simon, B.~Mercey, W.~Prellier, R.~Fr\'esard and P.~Limelette, \apl \textbf{91}, 262103 (2007).

\bibitem{mayr} M. Mayr, A. Moreo, J. A. Verg$\acute{e}$s, J. Arispe, A. Feiguin, and E. Dagotto, Phys. Rev. Lett \textbf{86}, 135 (2000).

\bibitem{DCR} G.~T.~Seidler, S.~A.~Solin and A.~C.~Marley, Phys. Rev. Lett. \textbf{76}, 3049 (1996).

\bibitem{merithew} R.~D.~Merithew, M.~B.~Weissman, F.~M.~Hess, P.~Spradling, E.~R.~Nowak, J.~O'Donnell, J.~N.~Eckstein, Y.~Tokura and Y.~Tomioka, Phys. Rev. Lett. \textbf{84}, 3442 (2000).

\bibitem{pala} A.~Palanisami, R.~D.~Merithew, M.~ B.~Weissman, Maitri P.~Warusawithana, F.~M.~Hess and J.~N.~Eckstein, Phys. Rev. B \textbf{66}, 092407 (2002).

\bibitem{perco} S. Kirkpatrick, Rev. Mod. Phys. 45, 574 (1973).

\bibitem{stress} D.~N.~Nichols, R.~J.~Sladek and H.~R.~Harrison, Phys. Rev. B \textbf{24}, 3025 (1981).

\bibitem{andres} A.~de Andrés, M.~García-Hernández, and J.~L.~Martínez, Phys. Rev. B \textbf{60}, 7328 (1999).

\bibitem{taran} S.~Taran, S.~Karmakar, S.~Chatterjee, B.~K.~Chaudhuri, C.~P.~Sun, C.~L.~Huang, and H.~D.~Yang, J. Appl. Phys. \textbf{99}, 073703 (2006).

\bibitem{finger80} L.~W.~Finger and R.~M.~Hazen, J. Appl. Phys. \textbf{51}, 5362 (1980).

\bibitem{luo2004}  Q.~Luo, Q.~Guo and E.~G.~Wang, Appl. Phys. lett. \textbf{84}, 2337 (2004). 



\end{thebibliography}
\end{document}